\documentclass[sigconf]{acmart}
\copyrightyear{2026}
\acmYear{2026}
\setcopyright{cc}
\setcctype{by}
\acmConference[SIGIR '26] {Proceedings of the 49th International ACM SIGIR Conference on Research and Development in Information Retrieval}{July 20--24, 2026}{Melbourne, Australia.}
\acmBooktitle{Proceedings of the 49th International ACM SIGIR Conference on Research and Development in Information Retrieval (SIGIR '26), July 20--24, 2026, Melbourne, Australia}
\acmISBN{979-8-4007-1592-1/2026/07}
\acmDOI{10.1145/XXXXXX}
\AtBeginDocument{%
  \providecommand\BibTeX{{%
    Bib\TeX}}}

\usepackage{amsmath,amsfonts}
\usepackage{algorithmic}
\usepackage{array}
\usepackage{balance}
\usepackage[caption=false,font=normalsize,labelfont=sf,textfont=sf]{subfig}
\usepackage{textcomp}
\usepackage{stfloats}
\usepackage{url}
\usepackage{cite}
\usepackage{verbatim}
\usepackage{graphicx}
\def\BibTeX{{\rm B\kern-.05em{\sc i\kern-.025em b}\kern-.08em
    T\kern-.1667em\lower.7ex\hbox{E}\kern-.125emX}}
\usepackage{nicematrix}
\usepackage{balance}
\usepackage{multirow}
\usepackage{enumerate}
\usepackage[utf8]{inputenc}
\usepackage{graphicx}
\usepackage{float}
\usepackage{hyperref}
\usepackage{adjustbox}
\usepackage{enumitem}
\usepackage{booktabs}
\usepackage{caption}
\usepackage{xcolor}
\usepackage[table]{xcolor}
\usepackage{algorithm}
\usepackage{algorithmic}
\usepackage{soul} 
\begin{document}

\author{Yu Wang}
\orcid{0009-0008-6272-8714}
\affiliation{%
  \institution{Hefei University of Technology}
  \city{Hefei}
  \country{China}}
\email{wangyu20001162@gmail.com}

\author{Yonghui Yang}
\authornote{Yonghui Yang is the corresponding author.}
\affiliation{%
  \institution{National University of Singapore}
  \city{Singapore}
  \country{Singapore}}
\email{yyh.hfut@gmail.com}

\author{Le Wu}
\affiliation{%
  \institution{Hefei University of Technology}
  \city{Hefei}
  \country{China}}
\email{lewu.ustc@gmail.com}

\author{Yi Zhang}
\affiliation{%
  \institution{Auhui University}
  \city{Hefei}
  \country{China}}
\email{zhangyi.ahu@gmail.com}

\author{Fei Liu}
\affiliation{%
  \institution{Hefei University of Technology}
  \city{Hefei}
  \country{China}}
\email{feiliu@mail.hfut.edu.cn}

\author{Richang Hong}
\affiliation{%
  \institution{Hefei University of Technology}
  \city{Hefei}
  \country{China}}
\email{hongrc.hfut@gmail.com}

\renewcommand{\shortauthors}{Yu Wang, Yonghui Yang, Le Wu, Yi Zhang, Fei Liu and Richang Hong.}

\newcommand{\fullname}{\textit{\textbf{Ha}rdness-aware and \textbf{No}ise-regularized preference optimization for \textbf{Rec}ommendation~(HaNoRec)}}
\newcommand{\shortname}{\textit{HaNoRec}}

\title{Multimodal Large Language Models with Adaptive Preference Optimization for Sequential Recommendation}
\begin{abstract}

Recent advances in Large Language Models (LLMs) have opened new avenues for sequential recommendation by enabling natural language reasoning over user behavior sequences. A common approach formulates recommendation as a language modeling task, where interaction histories are transformed into prompts and user preferences are learned via supervised fine-tuning. However, these methods operate solely in the textual modality and often miss users’ fine-grained interests, especially when shaped by rich visual signals such as product images or movie posters. Multimodal Large Language Models (MLLMs) offer a promising alternative by aligning text and vision in a shared semantic space. A prevalent training paradigm applies Supervised Fine-Tuning (SFT) followed by Direct Preference Optimization (DPO) to model user preferences. Yet, two core challenges remain: 1) Imbalanced sample hardness, where random negative sampling causes overfitting on easy examples and under-training on hard ones; 2) Cross-modal semantic bias, where the fixed reference model in DPO prevents the policy model from correcting modality misalignments—especially over long sequences. To address these issues, we propose a Multimodal LLM framework that integrates \fullname~. Specifically, \shortname~dynamically adjusts optimization weights based on both the estimated hardness of each training sample and the policy model’s real-time responsiveness, prioritizing harder examples. It further introduces Gaussian-perturbed distribution optimization on output logits to enhance cross-modal semantic consistency and reduce modality bias inherited from the reference model. 
Experiments on three benchmarks demonstrate its superiority over state-of-the-art methods. 
The code is available at http://github.com/wangyu0627/HaNoRec.
\end{abstract}

\maketitle

\section{Introduction}
Recently, Large Language Models (LLMs \citep{2017transformer, 2023llm_survey, 2024llama3}) have achieved significant progress in tasks such as natural language understanding \citep{2019bert}, semantic reasoning \citep{2022cot}, and knowledge generalization \citep{2024RAG}. This advancement motivates researchers to explore their potential applications in broader tasks \citep{2024bias, 2023multimodal, 2024llmgraph}. Especially in Recommender Systems (RS \citep{2010rs, 2017ncf, 2025IHGCL}), effectively capturing users' long-term preferences and dynamic needs remains a core challenge. Sequential Recommendation (SR \citep{2018SASRec, 2022CL4SRec, 2019bert4rec}) methods focus on mining users’ historical interaction sequences to infer the content they may be interested in next, and LLMs’ superior contextual comprehension and generative abilities offer new opportunities in SR. As shown in Figure \ref{fig:introduction} (a), the left side illustrates the LLM-based SR paradigm, where many studies \citep{2023tallrec, 2023llamarec, 2025bigrec} convert behavioral sequences into textual input prompts. For example, the user has watched the following movies: $[item_1]$, $[item_2]$, $\dots$, $[item_n]$. Please write a movie that the user may watch: $[candidate_1]$, $[candidate_2]$, $\dots$, $[candidate_k]$. However, LLMs typically focus solely on textual information and struggle to fully exploit the multimodal data \citep{2023MoRec} (e.g., images and video) that are prevalent in recommendation tasks, limiting their ability to understand rich and diverse recommendation contexts.

Some studies \citep{2024lcrec, 2025collm, 2023prompt} explore applying soft prompt strategies to multimodal information to alleviate the above limitations. For example, in Figure \ref{fig:introduction} (a), they use a frozen visual encoder to extract image features and project them into the LLM-based SR framework as tokens to improve recommendation performance. Although this strategy partially utilizes image information and yields performance improvements, the LLM itself lacks multimodal alignment training \citep{2024vlfeedback}, limiting its understanding of image semantics and deep connections between visual and textual data. In contrast, Multimodal Large Language Models (MLLMs \citep{2025Qwen2.5-VL, 2024llava, 2021clip}) successfully align multimodal information into a unified semantic space through joint pretraining tasks \citep{2025Qwen2.5-VL}, shared encoder architectures \citep{2024llava}, cross-modal contrast \citep{2021clip}, and large-scale data training \citep{2023gpt4v}. This capability not only enhances the model's understanding of multimodal data but also offers new possibilities for multimodal modeling in recommender systems. As shown in the bar chart in Figure \ref{fig:introduction} (a), MLLMs demonstrate higher hit ratios across all datasets, significantly outperforming LLMs and soft prompt methods, proving the advantages of MLLMs in multimodal recommendation.
Nevertheless, enabling MLLMs to effectively capture the evolution of user preferences in SR remains some significant challenges, as follow:

\begin{figure*}[t]
    \centering
    \includegraphics[width=\linewidth]{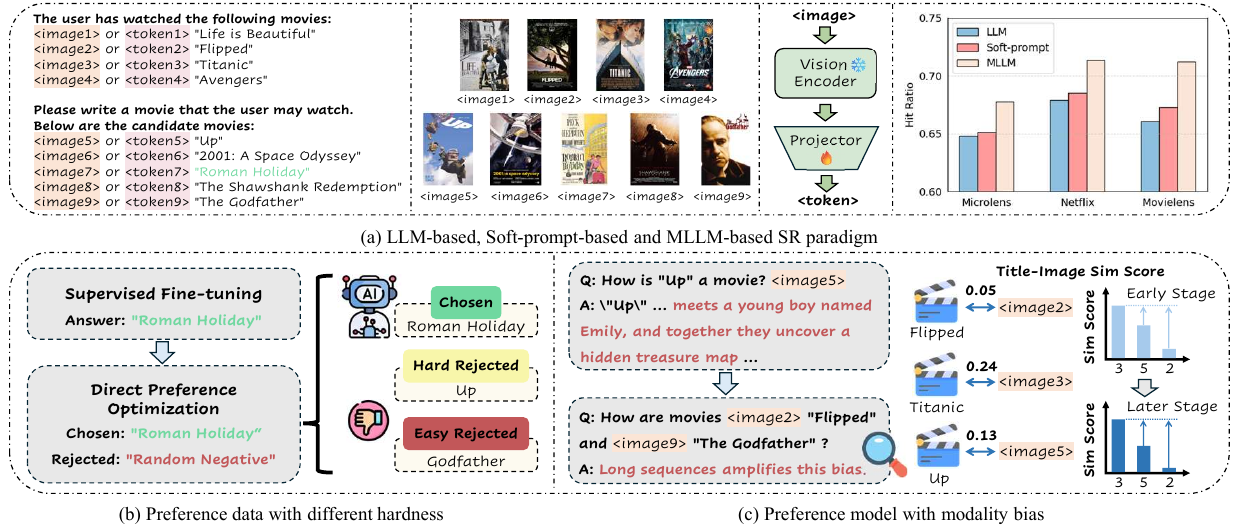}
    \caption{(a) Existing paradigms and performance comparisons of large model-based sequential recommendation; (b) Data construction and sample hardness issues in MLLM-SR during DPO training; (c) Title-image similarity scores exhibit a ``the weak get weaker'' trend due to MLLM modality bias after DPO training.}
    \label{fig:introduction}
    \vspace{-0.3cm}
\end{figure*}

\begin{itemize}[leftmargin=*]
\item \textbf{Imbalanced Sample Hardness.}
Existing MLLM-based SR methods \citep{2025MLLM-MSR, 2025MSRBench} generally adopt Supervised Fine-Tuning (SFT \citep{2023tallrec, 2025MLLM-MSR}), which focuses on observed historical behaviors while overlooking fine-grained changes in user preferences. Direct Preference Optimization (DPO \citep{2024rosepo, 2025SPRec}) further addresses this limitation by leveraging the differences between preference data. As shown in Figure \ref{fig:introduction} (b), SFT-based models construct DPO data by randomly sampling rejected examples from user interactions. However, this strategy results in noticeable imbalance across sample hardness levels, with the model favoring repeated training on easy contrasts and struggling to distinguish borderline cases or nuanced preferences. For example, an easy-to-distinguish sample like the ``\textit{Godfather}'' shows a large semantic gap with the positive sample ``\textit{Roman Holiday}'', while the gap is much smaller with ``\textit{Up}'', a semantically short and modality bias-prone sample. During DPO training, the reward gap continually amplifies the optimization of easy-to-distinguish samples while neglecting hard-to-distinguish ones, thus failing to fully capture hidden sequential preference signals. Therefore, designing a reweighting strategy that perceives variations in sample hardness is crucial.

\item \textbf{Cross-modal Semantic Bias.}
Nevertheless, a significant barrier in our alignment process is modality bias \citep{2023HA-DPO, 2024LLaVA-RLHF, 2024mdpo}, which occurs when the MLLM-based SR model fails to accurately interpret semantic information from an item's title and image, particularly in cases with ambiguous or short movie titles. As shown in Figure \ref{fig:introduction} (c), the model misdescribes ``Up'' (a film about adventure in a flying house) as ``searching for hidden treasure''. Furthermore, user histories often contain multiple such potentially ambiguous or thematically similar movies, and SR-based long sequences tend to amplify modality bias by accumulating these local semantic drifts \citep{2025MLLM-MSR, 2019sr_survey}. Accordingly, we calculate the title-image similarity scores throughout the alignment process, as illustrated in part (c). It can be observed that, samples with initially high similarity (e.g., ``Titanic'') remain stable or even slightly improve in the later stage. In contrast, samples with low initial similarity (such as ``Flipped'') show further score decline, indicating a “the weak get weaker” trend in representation degradation. Therefore, a sequence and distribution-sensitive optimization strategy for title-image similarity is needed.
\end{itemize}

To tackle the challenges mentioned above, we propose a preference optimization framework, named \fullname~. For \textbf{Challenge 1}, we design a \textbf{Hardness-aware Reweighting Strategy} (HaRS), where the dynamic scaling factor $\beta$ is proportional to sample hardness to mitigate sample imbalance and focus optimization on challenging cases. \textbf{In a nutshell, a larger factor constrains model updates, while a smaller factor encourages optimization \citep{2023dpo}}. Specifically, HaRS employs a specific MLLM encoder to convert multimodal information into semantic representations, calculates similarity scores to form positive-negative sample pairs for hardness estimation. It also evaluates model responsiveness through the normalized reward gap of each sample pair, and removes outliers.
For \textbf{Challenge 2}, we devise a \textbf{Noise-regularized Distribution Optimization} (NoDO) objective to mitigate cross-modal bias from ambiguous titles and maintain representational consistency in long sequences. NoDO introduces Gaussian noise perturbations to smoothly model the representation distributions of the current model and the reference model. It estimates their distributional distance in the title-image semantic space, and adaptively adjusts alignment constraints via KL divergence \citep{2016DVIB}.
The contributions are summarized as follows:

\begin{itemize}[leftmargin=*]
\item We investigate the application of MLLMs to sequential recommendation (SR) and design a hardness-aware reweighting strategy that effectively addresses sample imbalance and enhances the model’s sensitivity to subtle shifts in user preferences.
\item We further devise a noise-regularized distribution optimization objective, which aligns cross-modal title–image pairs via Gaussian noise perturbation and KL divergence-based constraints, thereby mitigating modality bias arising from weak semantics.
\item Extensive experiments on three benchmark datasets demonstrate that our approach consistently outperforms existing baselines. Furthermore, we provide a thorough analysis of the contributions of both the reweighting strategy and the distribution optimization method to SR performance.
\end{itemize}

\section{Preliminary}
In this section, we first introduce the problem of general sequential recommendation, the present how we integrate MLLMs to empower sequential recommendation through SFT and DPO manners. 

\noindent \textbf{Problem Formulation.}
In a general sequential recommendation scenario, we consider a user set $\mathcal{U}$ and an item set $\mathcal{V}$, where $|\mathcal{U}|$ and $|\mathcal{V}|$ denote the total number of users and items, respectively. The goal is to model dynamic user interests over time by leveraging their historical interaction behaviors. Formally, given an interaction sequence $S^u = \{v^u_1, v^u_2, \cdots, v^u_t\}$ for a user $u \in \mathcal{U}$, where $v^u_t$ is the item interacted with at timestamp $t$, and $t$ indicates the sequence length, the objective is to learn a recommendation function $f_\theta$ that predicts the next item $v^u_{t+1}$ the user is most likely to engage with.

\noindent \textbf{Supervised Fine-Tuning (SFT).}
The process of learning user preferences using an MLLM $\pi$ typically starts with an SFT phase, yielding a model denoted as $\pi_{\text{SFT}}$. Specifically, this stage involves adapting the pre-trained MLLM to multimodal tasks by fine-tuning it on a large-scale dataset $\mathcal{D} = \left\{x, \mathcal{I}, y\right\}$ composed of question–answer pairs, where $x$ is an instruction, $\mathcal{I}$ is the corresponding image input, and $y$ is the expected response generated by the model. Following the setting in \citep{2025MLLM-MSR}, each training sample $(x, \mathcal{I}, y)$ is built from the user’s interaction history. The instruction $x$ encodes the interaction sequence $S^u$ and a candidate set containing the ground-truth next item $v^u_{t+1}$. The image input $\mathcal{I}$ includes visuals of items in $S^u$ and $v^u_{t+1}$. The expected response $y$ is the ground-truth item $v^u_{t+1}$, which the model learns to generate or identify. The SFT objective aims to maximize the likelihood of generating the expected response $y$ conditioned on the instruction $x$ and the visual input $\mathcal{I}$. Formally, the training objective is:
\begin{equation}
\mathcal{L}_{\text{SFT}} = -\mathbb{E}_{(x, \mathcal{I}, y) \sim \mathcal{D}} \left[ \log\, \pi_{\text{SFT}}(y \mid \mathcal{I},x) \right].
\label{eq1}
\end{equation}

\noindent \textbf{Direct Preference Optimization (DPO).}  
Following Supervised Fine-Tuning (SFT), preference learning in MLLM-based recommendation is typically approached via \textit{Reinforcement Learning from Human Feedback (RLHF)}. In this setting, preference data are constructed as response pairs \((y_w, y_l)\), where \(y_w\) denotes the ground-truth next item and \(y_l\) is a randomly sampled unobserved item. Classical RLHF involves two stages: (1) training a reward model \(r_\phi(y \mid \mathcal{I}, x)\) to reflect user preference by scoring \(y_w\) higher than \(y_l\); and (2) optimizing the policy model \(\pi_\theta\) via reinforcement learning, such as Proximal Policy Optimization (PPO~\citep{2017ppo}), to align with the reward signal while remaining close to a reference model \(\pi_{\text{ref}}\):
\begin{equation}
\begin{aligned}
\max_{\pi_{\theta}} \ \mathbb{E}_{(\mathcal{I}, x) \sim \mathcal{D},\ y \sim \pi_{\theta}(\cdot \mid \mathcal{I}, x)} \left[ r^*_{\phi}(y \mid \mathcal{I}, x) \right] \\
- \beta \cdot D_{\text{KL}} \left[ \pi_{\theta}(y \mid \mathcal{I}, x) \parallel \pi_{\text{ref}}(y \mid \mathcal{I}, x) \right].
\end{aligned}
\label{eq2}
\end{equation}
To reduce the computational overhead of reward model training, \textit{Direct Preference Optimization (DPO)}~\citep{2023dpo} has been proposed as a more efficient alternative. Instead of learning an explicit reward function, DPO directly optimizes the target policy \(\pi_{\theta}\) using pairwise preference data, relying on implicit feedback embedded in the preference pairs. Its objective compares the log-likelihood ratios between preferred and rejected responses under both the policy and reference models:
\begin{equation}
\begin{split}
\mathcal{L}_{\text{DPO}} = - \mathbb{E}_{(\mathcal{I}, x, y_w, y_l)} [ \log \sigma \Big( &\beta \log \frac{\pi_\theta(y_w \mid \mathcal{I}, x)}{\pi_{\text{ref}}(y_w \mid \mathcal{I}, x)} \\
- &\beta \log \frac{\pi_\theta(y_l \mid \mathcal{I}, x)}{\pi_{\text{ref}}(y_l \mid \mathcal{I}, x)} \Big) ].
\end{split}
\label{eq3}
\end{equation}
Here, \(\beta\) is a hyperparameter controlling the trade-off between divergence regularization and preference alignment. \textbf{A larger \(\beta\) emphasizes staying close to the reference model by minimizing KL divergence, while a smaller \(\beta\) encourages stronger preference-driven updates~\citep{2023dpo, 2024mdpo}}. While effective, the standard DPO framework still overlooks key factors in multimodal recommendation, sample hardness, and cross-modal semantic misalignment. Next, we present our solution in detail.

\begin{figure*}[t]
    \centering
    \includegraphics[width=\linewidth]{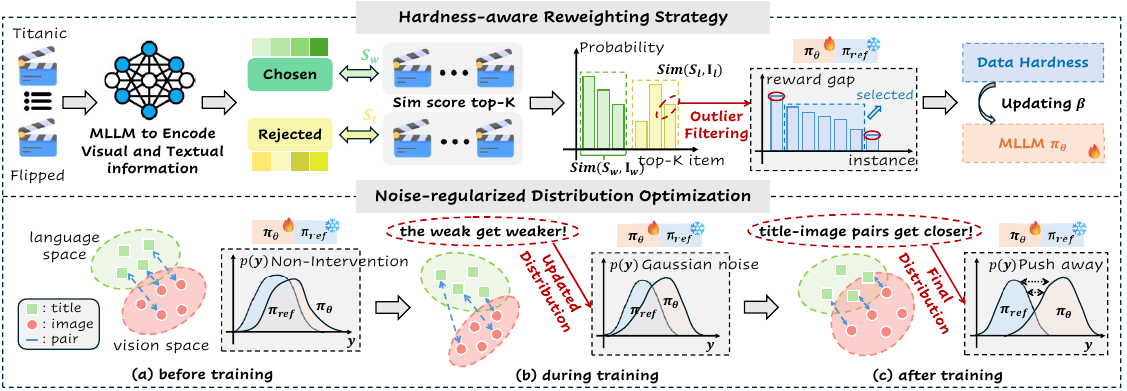}
    \caption{Illustration of \shortname. Hardness-aware Reweighting Strategy (HaRS) scales DPO weight proportionally to sample hardness, spotlighting challenging cases. Noise-regularized Distribution Optimization (NoDO) adds Gaussian noise and adaptive KL to align title–image semantics and curb modality bias.}
    \label{fig:model}
    \vspace{-0.3cm}
\end{figure*}

\section{Methodology}
Here, we introduce the technical implementation of our proposed \fullname~. As shown in Figure \ref{fig:model}, \shortname~ consists of two elaborated modules: Hardness-aware Reweighting Strategy~(HaRS) and Noise-regularized Distribution Optimization~(NoDO). Among them, HaRS dynamically adjusts the optimization emphasis for each training sample based on its hardness and the policy model’s responsiveness, ensuring that the model allocates more learning capacity to informative, hard-to-distinguish examples. NrDO, on the other hand, aims to alleviate modality bias by enhancing cross-modal semantic alignment. 
\subsection{Hardness-aware Reweighting Strategy} \label{sec3.1}
While DPO leverages preference pairs to optimize ranking performance, it relies on a fixed weighting scheme that may inadvertently reinforce easy samples while neglecting harder, more informative ones. To address this, we propose an adaptive weighting mechanism for the DPO scaling factor $\beta$ (refer to Eq.~\eqref{eq3}) that dynamically modulates the learning signal based on sample hardness and model responsiveness.
Formally, the training data are defined as $\mathcal{D} = \{(\mathcal{I}, x, y_w, y_l)\}$, where $\mathcal{I}$ denotes the image input, $x$ the instruction or query, and $y_w$, $y_l$ are the preferred and less preferred responses, respectively. Prior to training, we employ a pretrained MLLM~\citep{2025Qwen2.5-VL} to encode the item title set $\mathcal{V}$ and its corresponding image set $\mathcal{I}$ into semantic representations: a textual embedding matrix $\mathbf{H} \in \mathbb{R}^{|\mathcal{V}|\times d}$ and a visual embedding matrix $\mathbf{X} \in \mathbb{R}^{|\mathcal{I}|\times d}$, where $d$ is the shared embedding dimension.
To estimate sample hardness, we use the chosen ($y_w$) and rejected ($y_l$) items as anchors to compute their semantic similarity with all candidates in the embedding space. Specifically, we calculate similarity scores and retrieve the Top-K most similar items for each anchor, forming positive and negative sample clusters. These clusters serve as a proxy to quantify the contextual indistinguishability of each training pair—samples with overlapping clusters are considered harder and are thus assigned greater weight during optimization.
\begin{equation}
\mathcal{S}_* = \operatorname{Top\text{-}K}_{v_j \in \mathcal{V}} \left( \frac{(\mathbf{h}_j + \mathbf{x}_j)^\top (\mathbf{h}_{y_*} + \mathbf{x}_{y_*})} {\| \mathbf{h}_j + \mathbf{x}_j \|_2 \, \| \mathbf{h}_{y_*} + \mathbf{x}_{y_*} \|_2} \right), \quad * \in \{w, l\}
\label{eq4}
\end{equation}
where $(\mathbf{h}_j+\mathbf{x}_j)$ denotes the embedding of item $v_j$, and $(\mathbf{h}_{y_*} + \mathbf{x}_{y_*})$ represents the embedding of the corresponding target item. This sampling strategy captures multi-granularity preference signals, where varied positive samples reflect different aspects of preference (e.g., genre, style, visuals), and diverse negative samples provide gradients from easy to hard, enabling fine-grained ranking \citep{2024rosepo}. After obtaining sample sets $\mathcal{S}_w=\{\mathcal{S}_{w,1}, \mathcal{S}_{w,2}, \dots, \mathcal{S}_{w,K}\}$ and $\mathcal{S}_l=\{\mathcal{S}_{l,1}, \mathcal{S}_{l,2}, \dots, \mathcal{S}_{l,K}\}$, where $K$ denotes the number of $\operatorname{Top\text{-}K}$ similar items. Finally, we calculate $sim(\mathcal{S}_w, \mathcal{I}_w)$ and $sim(\mathcal{S}_l, \mathcal{I}_l)$ to estimate sample hardness and determine the dynamic $\beta$.
\begin{equation}
\mathbf{Z}_{*} = \operatorname{Softmax}(\big[sim(\mathcal{I}_{*}, \mathcal{S}_{*,k})\big]^{K}_{k=1}), * \in \{w,l\}
\label{eq5}
\end{equation}
where $sim(\cdot)$ denotes the cosine similarity function, and $\mathbf{Z}_w$ and $\mathbf{Z}_l$ represent the probabilities of preferred and rejected samples, respectively. In this step, we average the scores in $S_w$ to normalize sets of different sizes or ranges before applying Softmax function, avoiding dominance by set size or outliers and ensuring stable, differentiable probability distributions. The difference between the preferred and rejected probabilities indicates the data hardness, which can be defined as:
\begin{equation}
\Delta = \big\|\mathbf{Z}_{w,k} - \mathbf{Z}_{l,k}\big\|^{K}_{k=1}, \quad \lambda=\sigma(\Delta)/\sigma(\bar{\Delta})
\label{eq6}
\end{equation}
where $\Delta$ denotes the difference. The final data hardness $\lambda$ is computed by projecting $\Delta$ into the (0, 1) range using the sigmoid function $\sigma(\cdot)$, where $\bar{\Delta}$ is the average difference over the dataset. A larger hardness indicates an ``easy-to-distinguish'' instance, where the preferred sample aligns clearly with image-text semantics, increasing model confidence. A smaller hardness implies a ``hard-to-distinguish'' instance, where positive and negative samples are semantically or visually similar, making differentiation difficult and requiring a lower weight $\beta$ and more optimization attention.

However, $\lambda$ only distinguishes sample hardness at the data level and is calculated offline, which may misalign $\beta$ adjustment with the model’s actual responsiveness. Inspired by \citep{2025DAMO}, we compute the implicit reward gap to capture the model's current preference between positive and negative samples and guide the update of $\beta$. The reward gap is defined as:
\begin{equation}
\mathcal{R}_i = \left[ \beta \log \frac{\pi_{\theta}(y_{w,i} \mid x_i, \mathcal{I}_i)}{\pi_{\text{ref}}(y_{w,i} \mid x_i, \mathcal{I}_i)} - 
\beta \log \frac{\pi_{\theta}(y_{l,i} \mid x_i, \mathcal{I}_i)}{\pi_{\text{ref}}(y_{l,i} \mid x_i, \mathcal{I}_i)} \right],
\label{eq7}
\end{equation}
where $\pi_{\theta}$ and $\pi_{\text{ref}}$ represent the policy model and reference model, respectively. $\mathcal{R}_i$ corresponds to the reward gap of $\lambda_i$ for the $i$-th sample. Then, $\bar{\mathcal{R}}_i = \mathcal{R}_i/(\frac{1}{N}\sum_{i=1}^{N}\mathcal{R}_i)$($N$ denotes the number of samples) is mean-normalized, and a mask vector $\mathcal{M}_i=[0, 1, \cdots, 1, 0]$ filters out the two most extreme samples with the lowest and highest values to suppress outliers.
\begin{equation}
\eta = {\sigma\!\left(\dfrac{1}{N-2}\sum_{i=1}^{N} \mathcal{M}_i\,\bar{\mathcal{R}}_i\right)}/{\sigma(\bar{\mathcal{R}})},
\label{eq8}
\end{equation}
where $\eta$ is the estimated model responsiveness, and $\bar{\mathcal{R}}$ denotes data's estimated average reward gap. 
In other words, $\lambda$ measures the intrinsic difficulty of the data, while $\eta$ reflects the model's learning status. The model responsiveness allows $\beta$ to account for both sample difficulty and real-time training dynamics, compensating for limitations of similarity-based signals, and helps DPO avoid overfitting on easy samples and underfitting on hard ones, ensuring more stable and efficient optimization.

\subsection{Noise-regularized Distribution Optimization} \label{sec3.2}
In this section, we aim to mitigate the cross-modal semantic bias in MLLMs through the distribution optimization strategy of DPO. As shown in Figure \ref{fig:model} (a), before training, title embeddings (in language space) and image embeddings (in visual space) remain separated in the semantic space, increasing the risk of modality bias. At this point, the policy and reference models are untrained, and their distributions stay closely aligned.
Without intervention, DPO training leads to a “the weak get weaker” trend in Figure \ref{fig:model} (b). The output distributions of DPO show that, under the strong reverse KL constraint \citep{2025OPA-DPO}, policy model $\pi_\theta$ concentrates in high-probability regions of the reference model, further reducing similarity between misaligned title–image pairs. 

In such cases, the policy model behaves too similarly to the reference model, limiting its ability to overcome modality bias. To address this, we inject moderate Gaussian noise \citep{2024hallucination, 2025IRLLRec} into the output distributions of the policy model to increase modal flexibility.
\begin{equation}
\tilde{\pi}_{\theta}(y \mid x, \mathcal{I}) = \mathbb{E}_{\boldsymbol{\epsilon} \sim \mathcal{N}(0, \mathbf{I})} 
\left[ \pi_{\theta}(y \mid x, \mathcal{I}; \mathbf{z} + \boldsymbol{\epsilon} \odot \mathbf{z}) \right],
\label{eq9}
\end{equation}
where $\mathbf{z}$ denotes the last-layer logits of $\pi_{\theta}$ before generating the output distributions. Given the large number of base parameters and the use of LoRA fine-tuning \citep{2022lora}, we express the weights as $W = W_0 + BA$, where $A \in \mathbb{R}^{r \times d}$ and $B \in \mathbb{R}^{d \times r}$. During forward propagation, the LoRA increment is temporarily perturbed.
\begin{equation}
\begin{aligned}
A' &= A + \sigma_n \boldsymbol{\epsilon}_A, \quad \boldsymbol{\epsilon}_A \sim \mathcal{N}(0, \mathbf{I}) \\
B' &= B + \sigma_n \boldsymbol{\epsilon}_B, \quad \boldsymbol{\epsilon}_B \sim \mathcal{N}(0, \mathbf{I})
\end{aligned}
\label{eq10}
\end{equation}
and we computes the logits $W'= W_0 + B' A'$. This applies noise regularization to the trainable parts of $\pi_{\theta}$ and $\pi_{ref}$, resulting in naturally smoothed output distributions. As DPO training progresses through iterations, the Gaussian-smoothed distributions are incorporated into the similarity-aware adaptive KL divergence loss. Finally, as shown in Figure \ref{fig:model} (c), training leads to a more convergent semantic space with reduced modality bias. The joint optimization with Gaussian perturbation and adaptive KL regulation gradually aligns title and image embeddings toward a shared semantic center.

\subsection{Model Optimization} \label{sec3.3}
In a mini-batch, we define $\mathcal{B} = \left\{ (x_i, \mathcal{I}_i, y_{w,i}, y_{l,i}) \right\}_{i=1}^{B} \sim \mathcal{D}$ to analyze the optimization process during each epoch. We first calculate the data hardness for each sample pair, and then derive $\lambda_{\mathcal{B}}=\{\lambda_1,\lambda_2,\cdots,\lambda_{B}\}$ based on the batch partition $B$, and obtain the corresponding model responsiveness $\eta_{\mathcal{B}}=\{\eta_1,\eta_2,\cdots,\eta_{B}\}$. Next, we filter out outliers in $\eta_{\mathcal{B}}$ using Eq. \eqref{eq8} to obtain the batch-level model responsiveness $\bar{\eta}_{\mathcal{B}}$. We then compute the final weight for each sample through a dynamic combination method:
\begin{equation}
\beta^{'}_{\mathcal{B}} = \bar{\eta}_{\mathcal{B}} \cdot \lambda_{\mathcal{B}} \cdot \beta_0,
\label{eq11}
\end{equation}
where $\beta_0$ is the initial value, typically set to 0.1. This combination of HaRS enables the model to perceive data hardness while leveraging model responsiveness to enhance optimization robustness. 

Moreover, NoDO adaptively integrates Gaussian noise with implicit KL divergence \citep{2023dpo} in DPO to pull high-similarity pairs closer and push hallucinated pairs apart, reducing cross-modal semantic bias. Finally, the revised DPO loss is presented as loss $\mathcal{L}_{\text{HaNo}}$:
\begin{equation}
\begin{split}
\mathcal{L}_{\text{HaNo}} = - \mathbb{E}_{(\mathcal{I}, x, y_w, y_l)} [ \log \sigma \Big( &\beta^{'} \log \frac{\tilde{\pi}_{\theta}(y_w \mid \mathcal{I}, x)}{\pi_{\text{ref}}(y_w \mid \mathcal{I}, x)} \Big) \\
- &\beta^{'} \log \frac{\tilde{\pi}_{\theta}(y_l \mid \mathcal{I}, x)}{\pi_{\text{ref}}(y_l \mid \mathcal{I}, x)} \Big) ],
\end{split}
\label{eq12}
\end{equation}
where $\tilde{\pi}_{\theta}$ and $\beta^{'}$ represent the perturbed policy model and its corresponding adaptive dynamic DPO weight, respectively. The two modules collaborate to help the model focus on hard examples while preserving stable cross-modal alignment, addressing our proposed challenges in MLLM-based sequential recommendation. Algorithm~\ref{alg:HaNoRec} presents the complete training procedure of HaNoRec.

\subsection{Discussion of HaNoRec}
Direct impact on the DPO objective, which mitigates the “weak-gets-weaker” effect discussed in Section~\ref{sec3.2}. The DPO gradient, ignoring constants, is:
\begin{equation}
    \nabla L_{\mathrm{DPO}}(\theta) = - \mathbb{E}\bigl[\, \sigma(-\beta \Delta_\theta)\, \nabla \Delta_\theta \,\bigr], \Delta_\theta = f_\theta(x, y^{+}) - f_\theta(x, y^{-}).
    \label{eq13}
\end{equation}

When $|\Delta_\theta|$ is large (i.e., in the saturation regime), we have $\sigma(-\beta\Delta_\theta) \rightarrow 0$, leading to significant gradient decay, which is more severe for weak semantics or long sequences. Introducing zero-mean parameter noise $\delta$ (random smoothing) yields.
\begin{equation}
    \nabla L_{\mathrm{DPO}}(\theta) = - \mathbb{E}\bigl[\, \sigma(-\beta \Delta_{\theta+\delta})\, \nabla \Delta_{\theta+\delta} \,\bigr],
    \label{eq14}
\end{equation}
which redistributes probability mass toward the non-saturated region. As a result,
\begin{equation}
    \mathbb{E}\bigl[\, \sigma(-\beta \Delta_{\theta+\delta}) \,\bigr]\; > \;\sigma(-\beta\Delta_\theta),
    \label{eq15}
\end{equation}
thereby restoring effective gradients and preventing weak samples from being further marginalized. This behavior is consistent with the view that random smoothing preserves non-zero gradients even under strong constraints~\citep{2021sam}.

\noindent \textbf{Time complexity analysis.}
HaNoRec adds only two lightweight modules in training; inference remains identical to MLLM, with no extra latency or overhead. HaRS first preprocesses textual/visual tokens and then performs Top-K-style retrieval, keeping the cost negligible for each training step. NoDO injects light Gaussian perturbation into LoRA during forward pass only for distribution smoothing, with no extra forward/backward cost. Let text tokens = $T$, visual tokens = $V$, total $N = T + V$, batch size $B$, dimension size $d$. Standard DPO time complexity is: $\mathcal{O}\bigl(BL[4N^2 d + 4N d^2]\bigr)$, where $L$ is the number of Transformer layers. HaNoRec complexity: $\mathcal{O}\bigl(BL[4N^2 d + 4N d^2]\bigr) + \mathcal{O}(B) + \mathcal{O}(BN)$. If $L$ and $d$ are constants, both simplify to: $\mathcal{O}(4BN^2)$. Thus, under the same settings, runtime and memory overhead are minimal. \textbf{Result: HaNoRec maintains similar efficiency to DPO.}

\begin{algorithm}[h!]
\caption{Algorithm of HaNoRec}
\begin{algorithmic}[1]
\item[] \textbf{Input:} Given a preference dataset $\mathcal{D} = {(\mathcal{I}, x, y_w, y_l)} \sim \mathcal{D}$, where $\mathcal{I}$, $x$, $y_w$, and $y_l$ represent the image, question (including interaction history and candidate set), chosen response, and rejected response, respectively.
\item[] \textbf{Output:} The title of the next predicted item.
\STATE Initialize the policy model parameters $\pi_\theta$.
\FOR{\text{epoch} in \{1, 2, \ldots, $E$\}}
\FOR{$\mathcal{B} = \left\{ (x_i, \mathcal{I}_i, y_{w,i}, y_{l,i}) \right\}_{i=1}^{B} \sim \mathcal{D}$}
\STATE $\mathcal{S}_w \gets \operatorname{Top\text{-}K}\{y_w\}, \; \mathcal{S}_l \gets \operatorname{Top\text{-}K}\{y_l\}$;
\STATE obtain $\lambda_{\mathcal{B}}$ with $\{\mathcal{I}_w, \mathcal{S}_w\}, \; \{\mathcal{I}_l, \mathcal{S}_l\}$; \hfill $\triangleright$ {Equ~\eqref{eq4} $\rightarrow$ \eqref{eq6}}
\STATE obtain $\mathcal{R}_i$ with $y_{w,i}$ and $y_{l,i}$; \hfill $\triangleright$ Equ~\eqref{eq7}
\STATE obtain $\bar{\eta}_{\mathcal{B}}$ with $\mathcal{R}_i$; \hfill $\triangleright$ Equ~\eqref{eq8}
\STATE obtain $\tilde{\pi}_{\theta}$ with noise $\epsilon_A$ and $\epsilon_B$; \hfill $\triangleright$ {Equ~\eqref{eq9} $\rightarrow$ \eqref{eq10}}
\STATE $\lambda_{\mathcal{B}} \gets \{\lambda_1,\lambda_2,\cdots,\lambda_{B}\}$, $\bar{\eta}_{\mathcal{B}} \gets \{\eta_1,\eta_2,\cdots,\eta_{B}\}$;
\STATE obtain $\beta^{'}_{\mathcal{B}}$ with $\lambda_{\mathcal{B}}$ and $\eta_{\mathcal{B}}$; \hfill $\triangleright$ Equ~\eqref{eq11}
\STATE compute loss w.r.t. $\beta^{'}_{\mathcal{B}}$ and $\tilde{\pi}_{\theta}$; \hfill $\triangleright$ Equ~\eqref{eq12}
\STATE update the gradient and model $\pi_\theta$.
\ENDFOR
\ENDFOR
\\
\textbf{until} {The optimization is converged.}
\end{algorithmic}
\label{alg:HaNoRec}
\end{algorithm}

\begin{table}[t]
\captionsetup{justification=centering}
\caption{Statistics of the experimental datasets.}
\begin{adjustbox}{width=0.45\textwidth}
\begin{NiceTabular}{ccccc}
\toprule
\textbf{Dataset} & \textbf{\#Users} & \textbf{\#Items} & \textbf{\#Interactions} & \textbf{Density} \\
\midrule
Microlens  & 25,411 & 41,081 & 223,263   & $2.1 \times 10^{-4}$ \\
Netflix    & 13,187 & 17,366 & 68,933    & $3.0 \times 10^{-4}$ \\
Movielens  & 6,040  & 3,952  & 1,000,209 & $4.2 \times 10^{-2}$ \\
\bottomrule
\end{NiceTabular}
\end{adjustbox}
\label{table:dataset}
\end{table}

\section{Experiments}
In this section, we conduct extensive experiments and answer the following research questions:
\begin{itemize}[leftmargin=*]
\item \textbf{RQ1:} How does HaNoRec compare to the current state-of-the-art (SOTA) framework in terms of performance?
\item \textbf{RQ2:} Are the key components in our HaNoRec delivering the expected performance gains?
\item \textbf{RQ3:} What are the reasons for model's superior performance?
\item \textbf{RQ4:} How do different hyperparameters affect HaNoRec?
\end{itemize}

\subsection{Experiment Setup}
\subsubsection{\textbf{Datasets}}
We evaluate our model on three multimodal datasets. A statistical overview of all the datasets is presented in Table \ref{table:dataset}. 
\begin{itemize}[leftmargin=*]
\item \textbf{Microlens}~\citep{2023microlens} is a short-video recommendation dataset that includes titles, cover images, and video information. In this work, only titles and images are used for MLLM. The average number of user interactions is 8.79.
\item \textbf{Netflix}~\citep{2024llmrec} comes from the data collected on Kaggle by LLMRec and is a user-based movie recommendation dataset. Nonetheless, the dataset is highly sparse, with each user interacting with only 5.23 items on average.
\item \textbf{Moivelenes}~\citep{2024DHGPF} is a classic movie recommendation dataset, where each movie includes a title, year, and corresponding poster. It is very dense, with an average sequence length of 165.56.
\end{itemize}

\subsubsection{\textbf{Baselines}}
Our HaNoRec employs the Qwen as the backbone to serve as a sequential recommender. Therefore, we divide the baselines into traditional, LLM-based, and MLLM-based recommendation methods.

\noindent\textbf{Traditional RS.}
GRU4Rec \citep{2016GRU4Rec} uses GRU \citep{2014GRU} to construct a session-level sequential model for next-click prediction.
SASRec \citep{2018SASRec} models the user's historical behavior sequence using a self-attention mechanism.
LightGCN \citep{2020lightgcn} removes the feature transformation and nonlinear activations of GCN for light recommendation.
LATTICE \citep{2021LATTICE} constructs a latent item–item graph to enhance multimodal recommendation performance.
CL4SRec \citep{2022CL4SRec} is the first to introduce contrastive learning into SR.
MoRec \citep{2023MoRec} explores whether pure modality information can replace traditional item ID representations.
RecDCL \citep{2024RecDCL} applies both batch-wise and feature-wise contrastive learning to collaborative filtering.

\noindent\textbf{LLM-based RS.}
TallRec \citep{2023tallrec} proposes supervised fine-tuning of large language models to adapt them for recommendation tasks.
LLaRA \citep{2024llara} designs a recommendation assistant that combines collaborative filtering signals with language models.
SPRec \citep{2025SPRec} leverages a self-play mechanism in LLMs to de-bias and guide the model toward more balanced outputs.

\noindent\textbf{MLLM-based RS.}
MSRBench \citep{2025MSRBench} evaluates the capability of vision-language multimodal large models in SR.
MLLM-MSR \citep{2025MLLM-MSR} fine-tunes multimodal LLMs to specialize them for SR tasks.

\subsubsection{\textbf{Metrics}}
We evaluate model performance using AUC, HR@3, and NDCG@3. AUC is used to assess the binary decision of whether the user will like a target item. HR@3 and NDCG@3 are employed to measure the model’s Top-K ranking ability. AUC does not require a candidate set, while HR and NDCG use the prediction target along with 9 randomly sampled negative items to form a candidate set of 10. The size of the candidate item set is standardized across all methods for fair comparison.

\subsubsection{\textbf{Implementations}}
To ensure fairness, both our experiments and baselines are conducted on a Linux server equipped with 8 A5880 GPUs. We use Qwen-2.5-3B-Instruct\footnote{https://huggingface.co/Qwen/Qwen2.5-3B-Instruct} and Qwen-2.5-VL-3B-Instruct\footnote{https://huggingface.co/Qwen/Qwen2.5-VL-3B-Instruct} as the base models for the LLM-based and MLLM-based backbones, respectively. LLM-based models are trained for 10 epochs, and MLLM-based models for 5. During the training of HaNoRec, we consistently set the LoRA rank to 8, the learning rate to 1e-4, and the gradient accumulation steps to 8. Following the setup of \citep{2024llara}, all datasets are split into training, validation, and test sets in an 8:1:1 ratio based on the number of sequences. Due to the limited sequence length of the datasets, all models use only the most recent 6 interactions, with the last one serving as the prediction target. Furthermore, we use the same data for training and testing all traditional baseline methods.

\begin{table*}[t]
\captionsetup{justification=centering}
\caption{Recommendation performance comparisons on different datasets. Subscript $M$ denotes multimodal approaches. The superscript * indicates the Imprvement is statistically significant where the p-value is less than 0.05.}
\begin{adjustbox}{width=0.95\textwidth}
\begin{NiceTabular}[c]{c|l|c|ccc|ccc|ccc}
\toprule[1pt]
\midrule
& \multirow{2}{*}{\textbf{Model}} & \multirow{2}{*}{\textbf{Year}} & \multicolumn{3}{c|}{\textbf{Microlens}} & \multicolumn{3}{c|}{\textbf{Netflix}} & \multicolumn{3}{c}{\textbf{Movielens-1M}} \\
& & & AUC & HR@3 & NG@3 & AUC & HR@3 & NG@3 & AUC & HR@3 & NG@3 \\
\midrule
\multirow{7}{*}{\textbf{Traditional}}
\rowcolor{gray!5} &\textbf{GRU4Rec}     \citep{2016GRU4Rec}    & ICLR’16    & 0.6980 & 0.6052 & 0.4683 & 0.6347 & 0.5521 & 0.4779 & 0.6942 & 0.6374 & 0.4656 \\
\rowcolor{gray!5} &\textbf{SASRec}      \citep{2018SASRec}     & ICDM’18    & 0.7128 & 0.5970 & 0.4588 & 0.6470 & 0.5781 & 0.4986 & 0.7131 & 0.6467 & 0.4780 \\
\rowcolor{gray!5} &\textbf{LightGCN}    \citep{2020lightgcn}   & SIGIR’20   & 0.7035 & 0.6070 & 0.4669 & 0.6465 & 0.5625 & 0.4822 & 0.7025 & 0.6308 & 0.4710 \\
\rowcolor{gray!5} &$\textbf{LATTICE}_M$ \citep{2021LATTICE}    & MM’21      & 0.7247 & 0.6369 & 0.4875 & 0.6621 & 0.6042 & 0.5126 & 0.7164 & 0.6424 & 0.4672 \\
\rowcolor{gray!5} &\textbf{CL4SRec}     \citep{2022CL4SRec}    & ICDE’22    & 0.7320 & 0.6515 & 0.5102 & 0.6820 & 0.6250 & 0.5230 & 0.7240 & 0.6556 & 0.4804 \\
\rowcolor{gray!5} &$\textbf{MoRec}_M$   \citep{2023MoRec}      & SIGIR’23   & 0.7144 & 0.6294 & 0.4863 & 0.6784 & 0.6406 & 0.5326 & 0.6824 & 0.6159 & 0.4536 \\
\rowcolor{gray!5} &\textbf{RecDCL}      \citep{2024RecDCL}     & KDD’24     & 0.7202 & 0.6246 & 0.4790 & 0.6673 & 0.6163 & 0.5107 & 0.7223 & 0.6474 & 0.4752 \\
\rowcolor{gray!5} &$\textbf{AB-Rec}_M$  \citep{2025AB-Rec}     & KDD’25     & 0.7337 & 0.6520 & 0.5024 & 0.6769 & 0.6211 & 0.5204 & 0.7188 & 0.6509 & 0.4772 \\
\midrule
\multirow{3}{*}{\textbf{LLM-based}}
\rowcolor{green!5} &\textbf{TallRec}  \citep{2023tallrec}  & Recsys’23  & 0.7253 & 0.6480 & 0.4918 & 0.6927 & 0.6788 & 0.5533 & 0.7303 & 0.6606 & 0.4815 \\
\rowcolor{green!5} &\textbf{LLaRA}    \citep{2024llara}    & SIGIR‘24   & 0.7375 & 0.6504 & 0.4894 & 0.7178 & 0.6892 & 0.5592 & 0.7392 & 0.6705 & 0.4900 \\
\rowcolor{green!5} &\textbf{SPRec}    \citep{2025SPRec}    & WWW’25     & 0.7450 & 0.6682 & 0.4938 & 0.7378 & 0.7042 & 0.5663 & 0.7380 & 0.6856 & 0.4932 \\
\midrule
\multirow{4}{*}{\textbf{MLLM-based}}
\rowcolor{blue!5} &\textbf{MSRBench} \citep{2025MSRBench}  & WWW’25     & 0.7418 & 0.6650 & 0.4942 & 0.7201 & 0.7020 & 0.5654 & 0.7312 & 0.6846 & 0.4913 \\
\rowcolor{blue!5} &\textbf{MLLM-MSR} \citep{2025MLLM-MSR}  & AAAI’25    & \underline{0.7544} & \underline{0.6776} & \underline{0.5110} & \underline{0.7448} 
                                                                        & \underline{0.7135} & \underline{0.5754} & \underline{0.7497} & \underline{0.7123} & \underline{0.5031} \\
\cmidrule{2-12}
\rowcolor{blue!5} &\textbf{HaNoRec}(Ours)                 &            & \textbf{0.7852*} & \textbf{0.7360*} & \textbf{0.5673*} & \textbf{0.7710*} & \textbf{0.7483*} 
                                                                        & \textbf{0.6191*} & \textbf{0.7732*} & \textbf{0.7599*} & \textbf{0.5403*} \\
\rowcolor{blue!5} &\textbf{Improv. \%}                     &            & 4.12\% & 8.62\% & 11.02\% & 3.52\% & 4.88\% & 7.59\% & 3.13\% & 6.68\% & 7.39\% \\           
\midrule
\bottomrule[1pt]                       
\end{NiceTabular}
\end{adjustbox}
\label{tab:rec_result}
\end{table*}

\subsection{Performance Comparisons(RQ1)}
Table \ref{tab:rec_result} shows the performance improvements of HaNoRec on three public datasets. We further discuss interesting observations in the results and provide possible explanations. All experimental results are averaged over five runs.
\subsubsection{\textbf{Limitations of Traditional Methods.}}
Sequential and collaborative filtering models rely on abundant interactions, while our data remains highly sparse (even with truncation on Movielens). Although CL4SRec, the strongest among them, uses contrastive learning-based sequence augmentation to ease the loosely defined cold-start issue, it still falls behind the LLM-based method SPRec by 1.8\%, 8.2\%, and 1.9\% in AUC across the three datasets, highlighting the performance bottleneck in sparse scenarios.

\subsubsection{\textbf{Limitations of LLM-based Methods.}}
Existing LLM-based SR methods typically rely on supervised fine-tuning (SFT), which focuses on historical behaviors but overlooks fine-grained shifts in user preferences. SFT tends to overexpose popular items, reducing diversity and introducing bias. In addition, text-only LLMs cannot leverage multimodal information like images, making it hard to address modality bias from ambiguous titles \citep{2023llmrec_fairness}. For example, as shown in the green part of Table \ref{tab:rec_result}, all three LLM methods lag behind the MLLM-based MLLM-MSR by over 8\% in HR and NDCG on the sparsest dataset, Microlens, reinforcing our motivation.

\subsubsection{\textbf{Superiority of HaNoRec.}}
As shown in Table \ref{tab:rec_result}, our model outperforms all baselines across all metrics, with gains of 11.02\% on Microlens, 7.59\% on Netflix, and 7.39\% on Movielens, demonstrating its effectiveness in capturing evolving user preferences. Notably, the NG@3 metric improves significantly, indicating more diverse and relevant recommendations while maintaining ranking quality and reducing popularity bias.

Overall, under experimental setting \citep{2025SPRec}, performance shows a clear decline from MLLM-based to LLM-based to traditional methods. Traditional models rely on historical clicks and struggle with sparse interactions. LLM-based methods offer strong language reasoning but overlook visual signals, such as cover images that often serve as the first cue in real scenarios. In contrast, MLLMs capture fine-grained semantics like genre and emotion from item posters, leading to slower performance degradation. Growing evidence \citep{2024LLaVA-RLHF, 2024llava} also shows that MLLMs increasingly understand images well, supporting the promise of this research direction.

\begin{figure}[t]
    \centering
    \includegraphics[width=\linewidth]{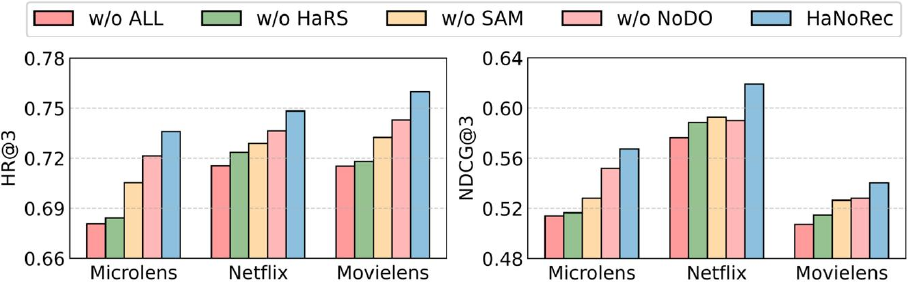}
    \caption{ Ablation studies of model variants on the all datasets w.r.t. HR@3 and NDCG@3.}
    \label{fig:ablation}
    \vspace{-0.5cm}
\end{figure}

\subsection{Ablation Study (RQ2)}
\subsubsection{\textbf{Impact of key components.}}
In this section, we assess the impact of key components in HaNoRec and provide potential explanations. The following model variants are evaluated:
\begin{itemize}[leftmargin=*]
\item HaNoRec w/o ALL: Removes all proposed components, only ``SFT + DPO'' training paradigm is conducted;
\item HaNoRec w/o HaRS: Removes the Hardness-aware reweighting strategy (HaRS) module in Sec. \ref{sec3.1};
\item HaNoRec w/o SAM: Removes the Top-K sampling strategy within the HaRS component;
\item HaNoRec w/o NoDO: Removes noise-regularized distribution optimization (NoDO) module in Sec. \ref{sec3.2}.
\end{itemize}
Figure \ref{fig:ablation} presents the experimental results on the all datasets, leading to the following insights:

Experimental results show that HaRS has the greatest impact on recommendation performance; without it, results often resemble those of the base model. This indicates that dynamically scaling the DPO weight effectively captures sample hardness and guides the model to focus on ``hard-to-distinguish'' samples, while assigning lower weights to ``easy-to-distinguish'' ones to avoid overfitting on simple patterns \citep{2025DAMO}. This is attributed to the fact that multiple positives capture diverse preference signals, while stratified negatives form a difficulty gradient that enhances ranking precision.

Subsequently, we further analyze the NoDO module. While its performance gain is modest, it addresses the modality bias in MLLM-based SR through a simple and effective Gaussian noise perturbation method. In a nutshell, all components improve accuracy when built upon ``SFT + DPO'' paradigm, highlighting the importance of preference optimization and distribution optimization in fully leveraging MLLMs' potential for sequential recommendation.

\subsubsection{\textbf{Impact of different MLLMs.}}
To evaluate the impact of different MLLMs as backbones, we run experiments with various Qwen-VL series and sizes, along with the widely used LLaVA-NeXT. As shown in Table \ref{table:backbone}, 2B models perform worst, while 7B versions of Qwen and LLaVA generally perform best, indicating that MLLM-based recommendation follows the large model scaling law \citep{2020scaling}. However, performance gains do not scale proportionally with time cost. The performance gap between LLaVA and Qwen may result from differences in their visual encoders and cross-modal alignment strategies. Overall, the generalization experiments confirm that our approach is model-agnostic and transferable, supporting its application to broader multimodal recommendation scenarios.

\begin{table*}[t]
\captionsetup{justification=centering}
\caption{Ablation studies of different MLLMs on three datasets w.r.t. AUC, HR@3 and NDCG@3.}
\begin{adjustbox}{width=0.75\textwidth}
\begin{NiceTabular}{c|ccc|ccc|ccc}
\toprule[1pt]
\multirow{2}{*}{\textbf{Variants}} & \multicolumn{3}{c}{\textbf{Microlens}} & \multicolumn{3}{c}{\textbf{Netflix}} & \multicolumn{3}{c}{\textbf{Movielens}} \\
& AUC  & H@3  & N@3 & AUC  & H@3  & N@3 & AUC  & H@3  & N@3 \\
\midrule
Qwen2-VL-2B   \citep{2024qwen2-VL}  & 0.7780 & 0.7338 & 0.5664 & 0.7724 & 0.7417 & 0.6130 & 0.7704 & 0.7486 & 0.5410 \\
Qwen2.5-VL-3B \citep{2025Qwen2.5-VL}& 0.7852 & 0.7360 & 0.5673 & 0.7710 & 0.7483 & 0.6191 & 0.7732 & 0.7599 & 0.5403 \\
Qwen2.5-VL-7B \citep{2025Qwen2.5-VL}& \underline{0.7890} & \underline{0.7421} & \underline{0.5756} & \underline{0.7816} & \textbf{0.7527} & \textbf{0.6238} & \underline{0.7785} 
                                    & \underline{0.7623} & \underline{0.5408} \\
LLaVA-NeXT-7B \citep{2024llava}     & \textbf{0.7957} & \textbf{0.7446} & \textbf{0.5805} & \textbf{0.7892} & \underline{0.7514} & \underline{0.6198} & \textbf{0.7831} & \textbf{0.7635} & \textbf{0.5422} \\
\bottomrule[1pt]
\end{NiceTabular}
\end{adjustbox}
\label{table:backbone}
\vspace{-0.2cm}
\end{table*}

\begin{figure}[t]
    \centering
    \includegraphics[width=\linewidth]{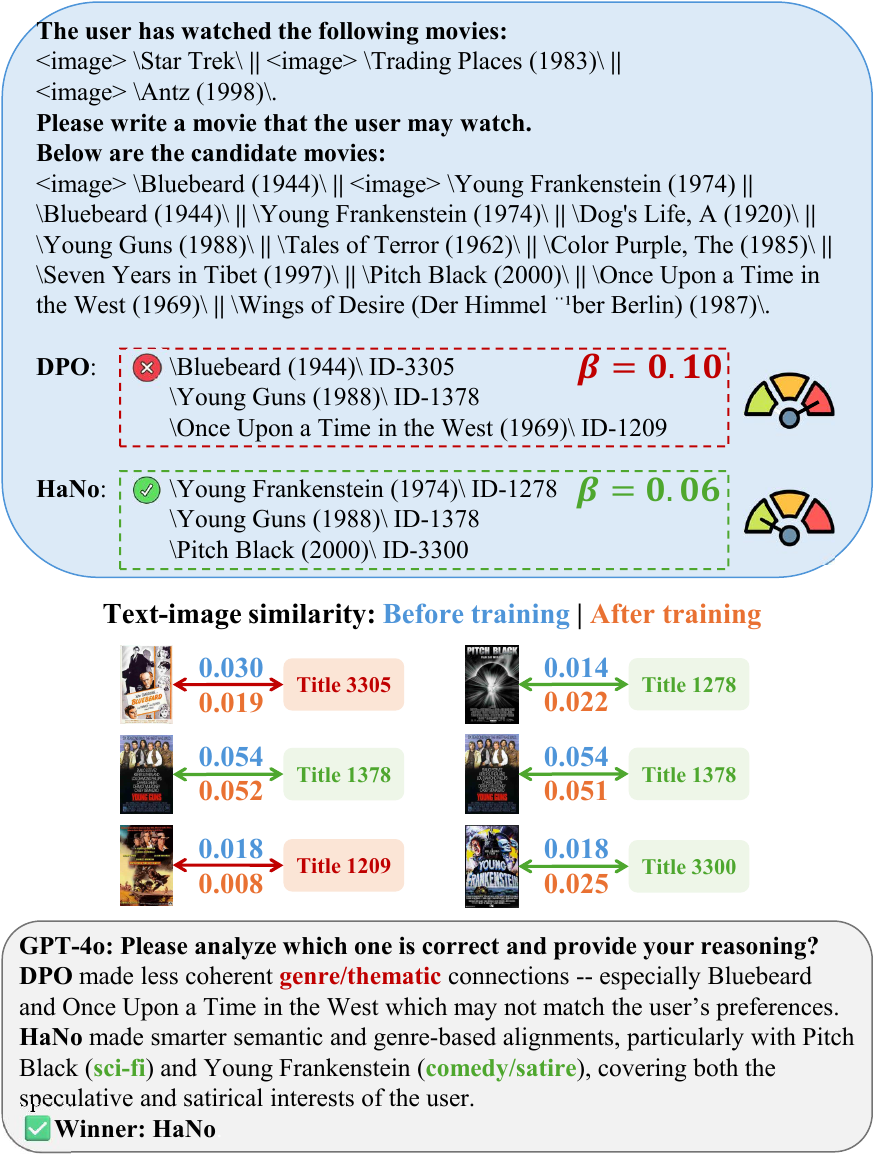}
    \caption{Case study on Movielens for user 4126.}
    \label{fig:case_study}
    \vspace{-0.5cm}
\end{figure}

\subsection{Case Study (RQ3)}
This section presents a case study of user 4126 from the Movielens dataset. As shown in the blue region of Figure \ref{fig:case_study}, given the interaction history and candidate set as a prompt, DPO produces an incorrect recommendation, while HaRS identifies the correct one. HaNoRec computes the sample’s hardness and model responsiveness, assigning a weight of 0.06, indicating it is hard to distinguish. HaRS relaxes the KL constraint and retrieves \textbf{Young Frankenstein}, which matches the user’s preference style. In contrast, DPO with fixed weighting recommends only popular but irrelevant items. 
Next, we compute the title–image similarity scores before and after training for the six movies retrieved by DPO and HaRS. Except for the top-scoring movie 1378, which slightly drops, DPO shows a significant decline in similarity for other movies, while HaRS improves the scores of its retrieved items. This supports \textbf{Challenge 2}, showing that NoDO effectively mitigates title–image mismatch caused by MLLM modality bias.

\begin{figure}[t]
    \centering
    \includegraphics[width=\linewidth]{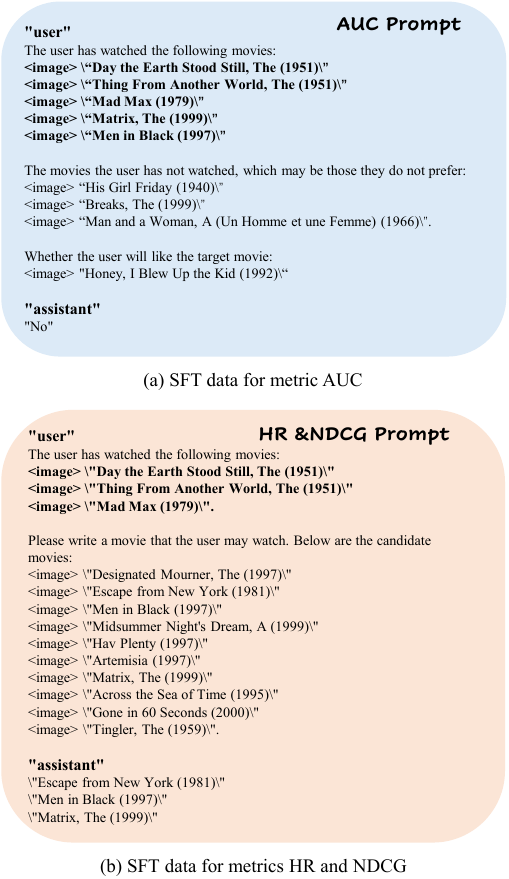}
    \caption{The prompt format for SFT paradigm.}
    \label{fig:prompt}
    \vspace{-0.5cm}
\end{figure}

Finally, we conduct a blind test using GPT‑4o \citep{2024gpt4o}, asking it to judge which recommendation list better matches the user based solely on historical interactions, without access to ground truth. As shown in the gray region of figure, GPT‑4o finds that HaRS’s recommendations better align with the user's interests in genre and semantics (sci-fi and dark comedy), while DPO’s list lacks thematic coherence, leading to the verdict: \textbf{Winner: HaNo}. In this section, we also introduce the automated construction process of the SFT and DPO datasets used by HaNoRec and baselines in Figure~\ref{fig:prompt}.

\begin{figure}[t]
    \centering
    \includegraphics[width=\linewidth]{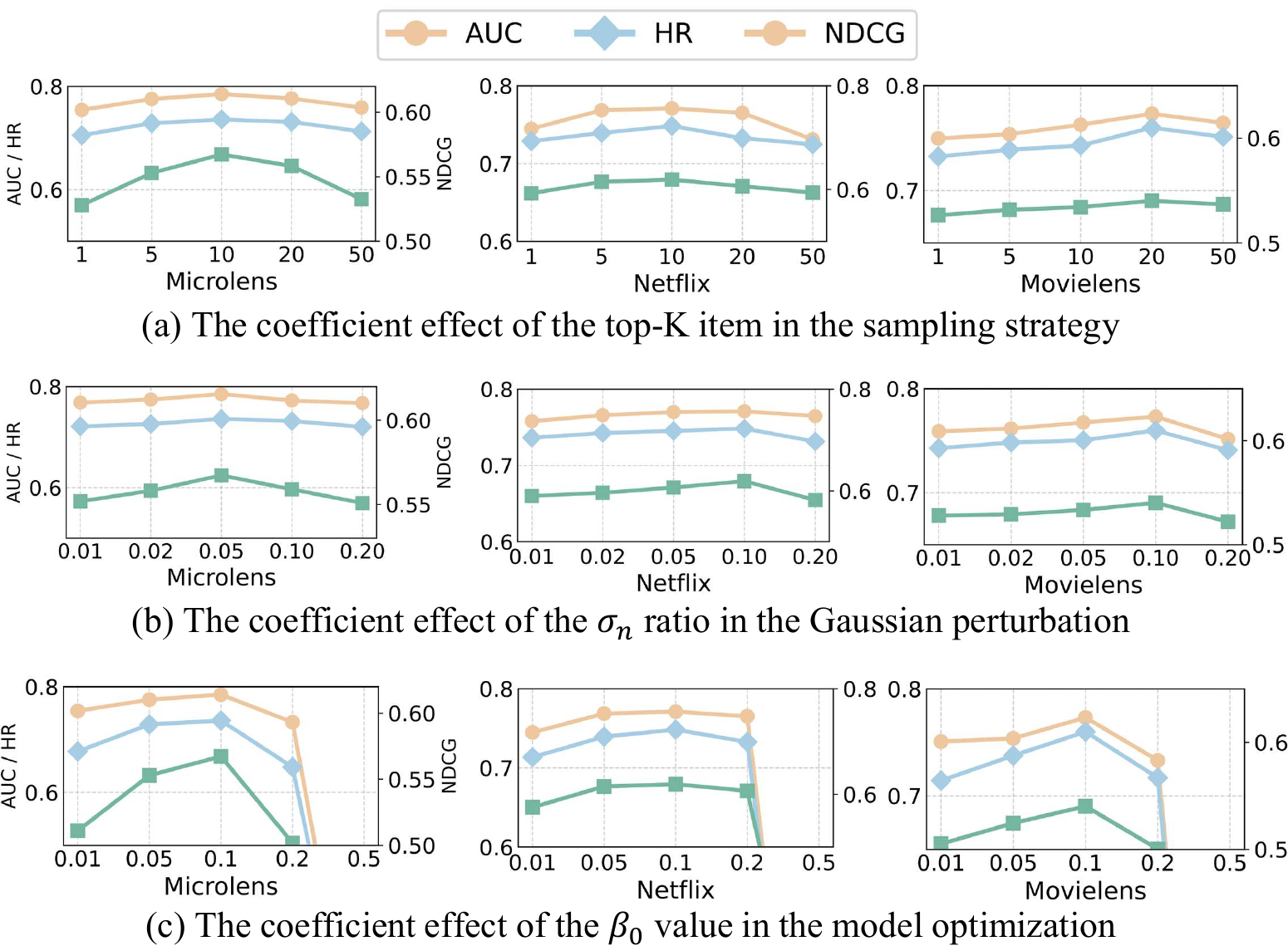}
    \caption{Performance comparison w.r.t. different hyperparameters. The $y$-axes on the left belong to the metrics AUC and HR and the right belong to NDCG.}
    \label{fig:hyperparameter}
    \vspace{-0.5cm}
\end{figure}

\subsection{Hyperparameter Sensitivity (RQ4)}
In this study, we keep HaNoRec parameters fixed and individually analyze the Top-K similar item group in HaRS, the Gaussian noise strength in NoDO, and the initial DPO weight, corresponding to $K$ in Eq. \eqref{eq5}, $\sigma_n$ in Eq. \eqref{eq10} and $\beta_0$ in Eq. \eqref{eq11}.
\begin{itemize}[leftmargin=*]
\item As shown in Figure \ref{fig:hyperparameter} (a), HaRS performs best when Top-K is set to 10 or 20. Too small (K=1) or too large (K=50) values lead to failure. With only one sample, it cannot capture multi-dimensional preferences; with too many weak or incorrect samples, the boundary between positive and negative clusters becomes unclear. In both cases, overly simple or noisy samples distort the hardness curve, making it hard for HaRS to focus on truly difficult instances and weakening optimization.
\item As observed from the figure, $\sigma_n$, the strength of Gaussian noise, has limited impact on performance. While 0.05 works best for Microlens, 0.1 is optimal for other datasets. A larger weight introduces stronger LoRA \citep{2022lora} perturbations, smoothing the distribution more—beneficial for datasets like Netflix with low-variance visual features. In contrast, a smaller weight provides milder perturbation and more stable distributions, which suits high-variance datasets like Microlens.
\item Finally, in part (c), we evaluate how different $\beta$ values affect the model. As the base coefficient increases, the model enlarges the positive–negative probability gap more rapidly, improving recommendation accuracy. When $\beta = 0.5$, performance drops significantly, as the DPO weight becomes too large, disrupting the fixed response pattern established during SFT.
\end{itemize}

\section{Related Work}
\noindent \textbf{LLMs and MLLMs for Recommendation.}
Large language models (LLMs) have gained significant attention in recommender systems for their advanced language understanding and reasoning abilities. Research in this area primarily follows three paradigms: LLMs as recommenders, enhancers, and encoders. LLMs as recommenders \citep{2024llara, 2023llamarec}: These methods use user interaction histories as prompts to guide LLMs in selecting recommendation targets from a candidate set. TallRec \citep{2023tallrec} fine-tunes Llama on constructed recommendation datasets, enhancing LLM decision-making in recommendations. RosePO \citep{2024rosepo} employs smoothing personalized preference optimization to fine-tune LLMs, improving performance while ensuring the recommender remains ``helpful and harmless''. LLMs as enhancers \citep{2025IRLLRec, 2024DALR, 2025DMRec}: RLMRec \citep{2024rlmrec} introduces a framework leveraging LLM-driven representation learning, with contrastive and generative alignment methods to improve recommendations. AlphaRec \citep{2024alpharec} replaces ID-based embeddings with language embeddings and combines GCN and CL for a simple yet effective recommendation approach. LLMs as encoders \citep{2023chatrec, 2024BLaIR}: EasyRec \citep{2024easyrec} leverages collaborative information and textual data from users and items to retrain language models for recommendation, achieving impressive performance in zero-shot scenarios. To extend LLM-based recommenders to multimodal scenarios, researchers explore multimodal large language models (MLLMs) that integrate visual and textual information to better capture user preferences. Rec-GPT4V~\citep{2024Rec-gpt4v} introduces a visual summarization reasoning mechanism to improve user modeling and reduce dynamic image redundancy. MSRBench~\citep{2025MSRBench} conducts a systematic evaluation of different MLLM integration strategies and highlights computational inefficiency as a key challenge. MLLM-MSR~\citep{2025MLLM-MSR} designs a two-stage user modeling framework that combines image summarization with sequence modeling to fine-tune MLLMs.

\noindent \textbf{Preference Alignment in MLLMs.}
Recent advances in large-scale human preference data \citep{2023human-dataset} and reinforcement learning (RL) techniques \citep{2022rlhf} have made preference alignment a key driver in the development of LLMs and MLLMs. In MLLMs, preference alignment research primarily focuses on two aspects: (I) constructing high-quality preference datasets \citep{2024vlfeedback, 2024RLHF-V} and (II) designing effective alignment optimization strategies \citep{2023dpo, 2024simpo}. Within the first direction, RLAIF-V \citep{2024RLHF-V} emphasizes a self-feedback alignment mechanism that spans both training and inference stages, while VLFeedback \citep{2024vlfeedback} focuses on large-scale automatically generated offline preference data. For alignment optimization strategies, the dominant methods are built upon the RL framework, adopting proximal policy optimization (PPO \citep{2017ppo}) or the simplified direct preference optimization (DPO \citep{2023dpo}) variant. For example, RLHF-RLAIF \citep{2024LLaVA-RLHF} improves reward design by incorporating vision-language factual enhancement and hallucination penalty mechanisms to enhance multimodal alignment. POVID \citep{2024POVID} generates AI preference data by injecting hallucinations and applying image perturbations, enabling feedback construction without human involvement. However, PPO-based methods face issues such as training complexity and poor stability \citep{2024mdpo}. Building on this, HA-DPO \citep{2023HA-DPO} enhances hallucination robustness by constructing consistent positive and negative sample pairs, while OPA-DPO \citep{2025OPA-DPO} alleviates this phenomenon by addressing the KL divergence in alignment strategy training. Subsequently, DAMO \citep{2025DAMO} dynamically adjusts the optimization strategy from both the data and model perspectives, revisiting the negative samples.

\section{Conclusion} 
This paper leveraged the capability of Multimodal Large Language Models (MLLMs) in understanding and generating multimodal information to improve sequential recommendation accuracy. We proposed a user preference learning framework, HaNoRec, which dynamically scaled the weight of the preference model based on hardness-aware reweighting strategy to focus on difficult samples and accurately capture the evolution of user preferences, while also incorporating model responsiveness to improve optimization robustness. Additionally, we divesed a noise-regularized distribution optimization objective to mitigate cross-modal bias via Gaussian noise perturbation and adaptive KL regularization. Experiments on three public datasets demonstrated the superiority of HaNoRec.


\bibliographystyle{ACM-Reference-Format}
\bibliography{sample-base}

\end{document}